\documentclass[prb,aps,twocolumn,showpacs]{revtex4}
\usepackage{txfonts}
\usepackage{amsfonts}
\usepackage{graphicx}
\usepackage{dcolumn}
\usepackage{bm}
\usepackage{color}

\begin{document}
\title{Impurity-induced bound states in superconductors with topological order}
\author{Fei Wang $^{1}$}
\author{Qin Liu $^{2}$ 
}
\author{Tianxing Ma $^{1,3,}$\footnote{Author to whom correspondence should be addressed. E-mail: txma@bnu.edu.cn}
}
\author{Xunya Jiang $^{2}$
}
\affiliation{$^1$ Department of Physics, Beijing Normal University,
Beijing 100875, China}
\affiliation{$^2$ State Key Laboratory of
Functional Materials for Informatics, Shanghai Institute of
Microsystem and Information Technology, CAS, Shanghai 200050, China}
\affiliation{$^3$ Beijing Computational Science Research Center, Beijing 100084, China}

\date{\today}

\begin{abstract}
The study of classical spins in topological insulators
[Phys. Rev. B {\bf 80}, 115216 (2009)] is generalized to topological
superconductors. Based on the characteristic features of the so-called
$F$-function, Bogoliubov-de Gennes Hamiltonian for superconductors is classified
to positive, negative, and zero ``gap'' categories for topologically
trivial and nontrivial phases of a topological superconductor as well as
a BCS superconductor respectively. It is found that the $F$-function
determines directly the presence or absence of localized excited
 states, induced by bulk classical spins and
nonmagnetic impurities, in superconducting gap and their persistence
with respect to impurity strength.  Our results provide an alternative
way to identify topologically insulating and superconducting phases
in experiments while without resorting to the surface properties.
\end{abstract}

\maketitle

\section{Introduction}
Time-reversal (TR) invariant topological insulators (TI) are new
state of matter in condensed matter physics, which have a full
insulating gap in the bulk, but gapless edge or surface states
consisting of an odd number of Dirac fermions. \cite{Review}
Materials for TR invariant TIs, including HgTe/CdTe quantum wells,
Bi$_{1-x}$Sb$_x$ alloy, binary (Bi$_2$Te$_3$, Sb$_2$Te$_3$,
Bi$_2$Se$_3$) and ternary (TlBiTe$_2$, TlBiSe$_2$) compounds, have
been confirmed experimentally according to the above characteristic
definition, by either transport measurements of the quantized edge
conductance $2e^2/h$, or angle-resolved photoemission spectroscopy
measurements of the linear dispersion of the surface states as well
as its odd numbers. \cite{Review} However, all these measurements
aim only at the surface properties of the TR invariant TIs.
While in a recent work \cite{Liu2009} by two of the authors, through
examining the localized excited states (LES) in the
insulating gap induced by bulk impurities in quantum anomalous Hall (QAH) systems, \cite{Qi2006}
in two-dimensional (2D) HgTe/CdTe quantum spin Hall (QSH) systems, \cite{Bernevig2006}
as well as in TR invariant 3D strong TIs, \cite{Fu2007} they point out that
distinctive behaviors between
topological and conventional insulators exist, which help to
distinguish topologically different insulating phases experimentally
without resorting to surface features.

Soon after the discovery of
TR invariant TIs, the study was generalized to TR invariant and
breaking topological superconductors (TSC).
\cite{Review,TSC,Schnyder2008} This generalization is natural
because there is a direct analogy between superconductors and
insulators where the Bogoliubov-de Gennes (BdG) Hamiltonian for the
quasiparticles of a superconductor is analogous to the Hamiltonian
of a band insulator, with the superconducting gap corresponding to
the band gap of the insulator. Instead of Dirac fermions, the
gapless surface states of a TR invariant TSC consist of odd number
of Majorana cones, which only have half the degree of freedom of Dirac
fermions. While for TR breaking TSC, the vortex core of which
carries an odd number of Majorana zero modes, \cite{Review} giving
rise to non-Abelian statistics and providing a possible platform for
topological quantum computing. \cite{Nayak2008} Several ways to
realize topological superconductivity by making use of the
superconducting proximity effect on the 2D surface
states of 3D TR invariant TIs, \cite{Fu2008} or on the 2D TR
breaking TIs, \cite{Qi2010} and on semiconductors with strong Rashba
spin-orbit coupling \cite{Sau2010} have been imposed. However, no
definitive proofs in experiments have yet been found so far.

Gaped systems can be classified into ten symmetry classes,
among which four are the BdG classes for superconductors. \cite{TSC,Schnyder2008}
Bearing in mind the extreme analogy between the BdG Hamiltonian of a
superconductor and that of a band insulator, we show in this work
that the BdG Hamiltonian for superconductors can also be assorted in another viewpoint
into three categories of positive, negative, and zero ``gap'' superconductors
according to the so-called $F$-functions. \cite{Shiba1968,Liu2009}
By doping into bulk magnetic (which we treat as classical spins \cite{classicalspin}) and
nonmagnetic impurities, we show that the characteristic properties
of the $F$-functions determine directly the presence or absence of
LES, induced by the impurities, in superconducting or insulating gap,
as well as the persistence of the LES with respect to impurity strength.
Based on this observation, a generic method is proposed,
by testing the response to classical spins as well as nonmagnetic
impurities in the bulk, to differentiate the topologically
nontrivial superconducting or insulating phases from the trivial
ones in general, and also to distinguish a TSC from a conventional BCS superconductor
in particular.

Specifically speaking, using the T-matrix method, \cite{Shiba1968} it is found that,
similarly as in TIs, \cite{Liu2009} in TSCs there are always
{\it four} LES in the superconducting gap
for spin-dependent potential, whereas only {\it two} such LES for
ordinary potential. Moreover, these LES survive under arbitrary
impurity strength. While when the TSC transits into the
topologically trivial phase, the LES exist only at small impurity
strength but then disappear into bulk bands at strong impurity
strength. The classical spins localized in a BCS superconductor have
been discussed by H. Shiba in 1968. \cite{Shiba1968} In the viewpoint of
$F$-functions, BCS superconductors fall into the critical category between
the topologically trivial and nontrivial phases of a TSC. It is shown
that {\it two} LES appear only for
spin-dependent potential but no such LES exist for ordinary
potentials. Moreover, these two LES persistently tend nearer to the band edges at strong impurity
strength, which is in a tricky contrast to the trivial phase of a TSC.
Therefore, through the observation of the LES in
superconducting (band) gap induced by bulk impurities, the potential
strength of both spin-dependent and ordinary impurities can be used
to tell the topologically nontrivial from the trivial phases of a
TSC (TI), whereas nonmagnetic impurities can be used to distinguish
a TSC from an ordinary (BCS) superconductor.

The rest of this paper is organized as the following. In Section
\ref{review}, we first briefly review the symmetry classification of
BdG Hamiltonian for superconductors. \cite{Schnyder2008} Through
this classification, all categories of BdG Hamiltonian can be
reduced, if the additional particle-hole symmetry (PHS) is imposed,
to the model Hamiltonian of either the 2D QSH system, \cite{Bernevig2006} or the 3D
strong TI, \cite{Fu2007} or half the spin degree of freedom of the
above two which breaks the TR symmetry. \cite{Qi2006,Read2000} The
responses to classical spins and nonmagnetic impurities of all these
correspondences in TIs have been studied in Ref.\cite{Liu2009}, with
the only exception being class C superconductors. \cite{Senthil1999}
In Section \ref{dwave}, we study the classical spins in
$d_{x^2-y^2}+id_{xy}$ superconductors as an specific example of
class C. It is shown that its response to both spin-dependent and
ordinary potentials obey the same rules as those \cite{Liu2009} in
TIs. Therefore, observing the LES in superconducting (band) gap
induced by bulk impurities serves as an effective criterion to tell
the topological superconducting (insulating) phase. This work is
finally concluded in Section \ref{conclusion}.
\begin{figure}
\begin{center}
\includegraphics[width=9cm]{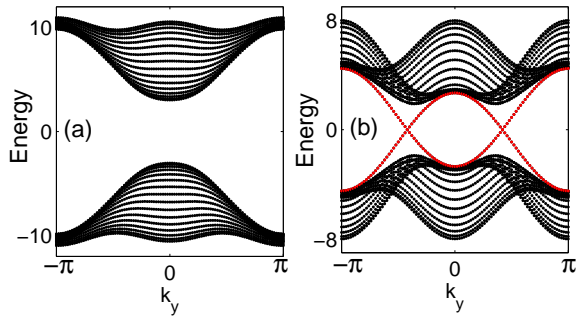}
\caption{(Color online) Energy spectra of $d+id$ superconductors
with open boundary condition in $x$-direction. Parameters
are taken as $\Delta_{x^2-y^2}=2$, $\Delta_{xy}=-2$, and all
energies are in unit of $t$. (a) Spectrum for a topologically trivial phase at
$\mu=-3$. (b) Spectrum for a topologically nontrivial phase at $\mu=3$. Two
gapless edge modes in the superconducting gap are marked in red.}
\label{spectrum}
\end{center}
\end{figure}

\section{Brief review of symmetry classification of BdG Hamiltonian
for superconductors} \label{review}

In this section, we briefly review the symmetry classification of BdG
Hamiltonian of superconductors. It is concluded that within all topologically
nontrivial categories of the four BdG classes in dimension two and three, \cite{note}
only class C in dimension two is left unaddressed for a complete discussion.

In a recent pioneering work by Schnyder {\it et al.},
\cite{Schnyder2008} ten symmetry classes of single-particle
Hamiltonians for gapped systems are classified according to the
presence or absence of TR symmetry, PHS, as well as sublattice
symmetry. Among which, four symmetry classes (named as D, DIII, C,
CI) arise in the BdG Hamiltonian for superconductors because of the
definitive PHS but the alternative presence of TR symmetry or SU(2)
spin rotational symmetry.

A general form of BdG Hamiltonian for the dynamics of quasiparticles
deep inside the superconducting state of a superconductor can be
written as \cite{Schnyder2008,Sigrist1991}
\begin{eqnarray}
H=\frac{1}{2}\sum_k\left(\begin{array}{cc}{\bf c}_k^{\dagger}&{\bf
c}_k\end{array}\right)\left(\begin{array}{cc}\varepsilon_k&
\Delta_k\\ \Delta_k^\dagger & -\varepsilon_{-k}^{T}
\end{array}\right) \left(\begin{array}{cc}{\bf c}_k\\{\bf c}_k^{\dagger}\end{array}\right),
\end{eqnarray}
where ${\bf c}_k$ and ${\bf c}_k^{\dagger}$ can be either column or
row vectors, $\varepsilon_k=k^2/2m-\mu$ is the
single-particle energy dispersion, $\Delta_k=({\bf d}_k\cdot
\mbox{\boldmath$\sigma$})(i\sigma_y)$ is the gap parameter with \mbox{\boldmath$\sigma$}
being the electron spin Pauli matrix vector, and ${\bf d}_k$ is a 3D
vector in spin space as a function of momentum $k$.

In class D, where neither SU(2) invariance nor TR symmetry is
present, there exist topologically nontrivial superconducting phases
in both 1D and 2D. A typical example of BdG Hamiltonian in this
class in 2D is the spinless chiral $p$-wave ($p\pm ip$)
superconductors, where the gap parameter is explicitly
$\Delta_k=\bar\Delta(k_x-ik_y)$, $\bar\Delta\in\mathbb{R}$. We
notice that this BdG Hamiltonian of TR breaking $p$-wave
superconductors is nothing else but the BHZ model
\cite{Bernevig2006} of a QSH system with half the spin degree of
freedom, or the QAH system studied by Qi
{\it et al.}, \cite{Qi2006} by dropping the terms proportional to
the identity matrix to maintain the generic PHS. In this sense, a
TSC can be viewed as a TI with PHS. Classical spins in QSH and QAH
systems have been studied in the context of TI, \cite{Liu2009} where
the term which breaks the PHS is shown to be unimportant to the
topological properties of the systems under study. Therefore it is
safe to conclude that the $p+ip$ superconductors have nontrivial
response to spin-dependent (ordinary) potential, where there are
always two (one) LES in the superconducting gap in topologically
nontrivial phase, which disappear in trivial phase.

In class DIII, though full SU(2) invariance is still absent but TR
symmetry is restored. The simplest way to regain TR symmetry is to
make two copies of the TR breaking system but with the whole being a TR conjugate pair. In this
class, topologically nontrivial superconducting phases exist in all
three dimensions. An interesting example in 2D is the equal
superposition of two chiral $p$-wave superconductors with opposite
chiralities ($p_x+ip_y$ and $p_x-ip_y$ waves), namely the helical
$p+ip$ superconductors, where the d-vector is explicitly ${\bf
d}_k=\bar\Delta(-k_x,k_y,0)$. In the basis of $({\bf
c}_{k}^{\dagger}\;{\bf
c}_{-k})=(c_{k\uparrow}^{\dagger}\;c_{-k\uparrow}\;c_{k\downarrow}^{\dagger}\;c_{-k\downarrow})$,
we see that the BdG Hamiltonian of the helical $p+ip$
superconductors is completely identical to that of the BHZ model in
HgTe/CdTe quantum wells with PHS, where spin up (down) electrons
form $p_x+ip_y$ ($p_x-ip_y$) Cooper pairs. Hence the response of
such helical $p+ip$ superconductors is similar to that of the QSH
systems, where persistent LES exist in the superconducting gap. In 3D,
a member of class DIII is the Balian-Werthamer (BW) state
\cite{Balian1963} of the B phase of liquid $^3$He described by the
d-vector ${\bf d}_k=\bar\Delta(k_x,k_y,k_z)$. In the same basis as
the helical $p+ip$ superconductors, the model Hamiltonian of the BW
state is the same as that of a 3D TI, which reduces exactly to the
2D QSH system in the limit $k_z\sim \langle k_z\rangle\simeq 0$. We
have also studied the $sp$-$d$ exchange coupling in 3D TIs,
\cite{Liu2009} again the LES behave distinctively in topologically
nontrivial and trivial phases, which is also valid to the BW state
of topological superfluids.

With the presence of full SU(2) invariance, the BdG Hamiltonians
fall into C and CI classes in the absence and presence of TR
symmetry. Nontrivial topological phases exist in 2D in class C,
while in 3D in class CI. A full focus on the five categories of TR
invariant TIs in 3D out of ten symmetry classes have been laid on in
the pioneering work by Schnyder {\it et al.}, \cite{Schnyder2008}
therefore for a complete discussion the only unaddressed one is the
2D case in class C, which is our main attention in this work. The interesting
example in class C in 2D is the TR breaking superconductors with
($d+id$)-pairing. \cite{Senthil1999} Under the basis $({\bf
c}_{k}^{\dagger}\;{\bf
c}_{-k})=(c^{\dagger}_{k\uparrow}\;c_{-k\downarrow})$, the gap
parameter in $d+id$ superconductors is
\begin{eqnarray}
\Delta_k=\Delta_{x^2-y^2}(k_x^2-k_y^2)+i\Delta_{xy}k_xk_y,
\label{delta}
\end{eqnarray}
where $\Delta_{x^2-y^2}$ and $\Delta_{xy}$ are real amplitudes.
\begin{figure}
\begin{center}
\includegraphics[width=7cm]{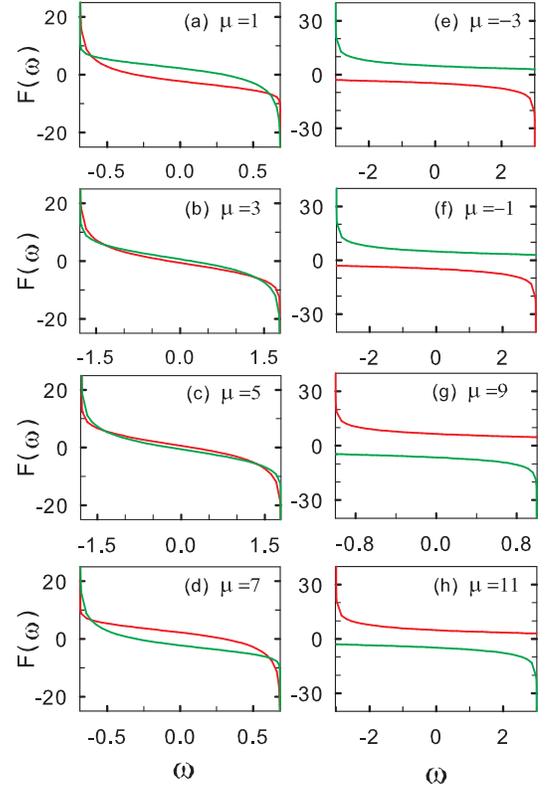}
\caption{(Color online) $F$-functions versus $\omega$ which lies in
bulk superconducting gap for different $\mu$'s. (a)-(d) $F$-functions
in topologically nontrivial phase, (e)-(h) $F$-functions in
topologically trivial phase. The electron band $F$-function
$F_e(\omega)$ is shown in red lines while the hole band $F$-function
$F_h(\omega)$ is shown as green lines. Parameters are taken as ${\rm
sgn}(\Delta_{x^2-y^2}\cdot\Delta_{xy})=-1$ and all energies are
measured in units of $t$.} \label{Ffunctionfig}
\end{center}
\end{figure}

In the following, we study the topological properties of the $d+id$
superconductors as well as its response to classical spins and
nonmagnetic impurities in bulk using the T-matrix method.
It is found that the $d+id$ superconductors show
the same behavior as that of TIs. Specifically, in
topologically nontrivial phase, there are four LES in the
superconducting gap for magnetic impurities and two LES for
nonmagnetic impurities at arbitrary impurity strength. These LES
then disappear into the bulk bands at large impurity strength as the
system transits into the topologically trivial phase. For
comparison, no LES appear with nonmagnetic impurities in a BCS
superconductor. Therefore, by testing the response to both magnetic
and nonmagnetic impurities, not only a topologically nontrivial
phase can be distinguished from a trivial one, but also a
conventional BCS superconductor can be discriminated from a TSC.

\section {Classical spins in $d_{x^2-y^2}+id_{xy}$ superconductors}
\label{dwave}

In this section, we study $d+id$ superconductors described by
Eq.(\ref{delta}) as an specific example in class C in parallel to
the previous work in Ref.\cite{Liu2009}. The results are stated
in the language of superconductors in particular, which are also
valid for insulating systems in general.

The single-particle
Hamiltonian of $d+id$ superconductors in momentum space is written as $h({\bf k})=f_{\alpha}({\bf
k})\sigma^{\alpha}$, where $\alpha=1,2,3$, and in tight-binding
model, ${\bf f}({\bf k})=(-\Delta_{x^2-y^2}(\cos k_x-\cos k_y),
\;-\Delta_{xy}\sin k_x\sin k_y,\;t(4-\cos k_x-\cos k_y)-\mu)$, with
$t$ being the hopping energy. The Hall conductivity of
this two-band system at zero temperature when the chemical potential
lies inside the band gap is calculated using the standard Kubo
formula as \cite{Qi2006}
\begin{eqnarray}
\sigma_{xy}=-\frac{1}{8\pi^2}\iint_{\rm FBZ}dk_xdk_y\hat{\bf
f}\cdot\partial_{k_x}\hat{\bf f}\times\partial_{k_y}\hat{\bf f},
\end{eqnarray}
where $\hat f_{\alpha}({\bf k})=f_{\alpha}({\bf k})/f({\bf k})$ is
the unit vector along the direction $f_{\alpha}$ and $f({\bf
k})=\sqrt{f_{\alpha}({\bf k})f^{\alpha}({\bf k})}$. This Hall
conductivity is related with the so-called Skyrmion number
\cite{Rajaraman,Qi2006} by $\sigma_{xy}=-Q_{\rm sky}/2\pi$, which is
an integer, and in $d+id$ superconductors it is shown to be
\begin{eqnarray}
Q_{\rm sky}=\left\{\begin{array}{cc}-2{\rm
sgn}(\Delta_{x^2-y^2}/\Delta_{xy}),&0<\mu<8,\\
0, &\mu<0\;{\rm or}\;\mu>8.\end{array}\right.\label{skymion}
\end{eqnarray}
Therefore the bulk-edge correspondences tell us that there should localize two
edge states at each boundary in the topologically nontrivial phase, which is indeed true as seen
in Fig.\ref{spectrum} where typical energy spectra of $d+id$
superconductors are shown.

To consider the single-particle excitations in the superconducting
gap, our starting point is to rewrite the BdG Hamiltonian of $d+id$
superconductors in Nambu space \cite{Shiba1968,Maki1967} as
\begin{eqnarray}
H_{\rm BdG}=\frac{1}{2}\sum_k
A^{\dagger}_k\left(\begin{array}{cccc}\varepsilon_k&\Delta_k & &\\
\Delta_k^{\ast}&-\varepsilon_k&&\\&
&\varepsilon_k&-\Delta_k\\&&-\Delta_k^{\ast}&-\varepsilon_k\end{array}\right)A_k,
\label{d+id}
\end{eqnarray}
where the gap parameter $\Delta_k$ is given in Eq.(\ref{delta}), and
$A_k^{\dagger}=(c_{k\uparrow}^{\dagger}\;c_{-k\downarrow}\;c_{k\downarrow}^{\dagger}\;c_{-k\uparrow})$.
This is because the Green's function (GF) formulation in Nambu space
can be easily generalized to include the paramagnetic impurities as
well as the Kondo effect. Also notice that the above BdG Hamiltonian of
a $d+id$ superconductor in Nambu space is formally similar to the 2D
Luttinger model in the hole bands of a semiconductor. \cite{Qi2006} In the presence of a localized spin or an
ordinary potential in bulk, their interactions between conduction
electrons are respectively
\begin{eqnarray}
H_{\rm ex}=(J/2)\sum_{kk^{\prime}}{\bf c}_k^{\dagger}{\bf S}\;{\bf
c}_{k^{\prime}}\cdot \mbox{\boldmath$\sigma$},\;\;
H^{\prime}=(V/2)\sum_{kk^{\prime}}{\bf c}_k^{\dagger}{\bf
c}_{k^{\prime}},
\end{eqnarray}
where ${\bf S}=(S_x,S_y,S_z)$ is the vector of a localized spin with
modulus $S^2=S_{\alpha}S^{\alpha}$. Again in Nambu space they take
the form
\begin{eqnarray}
&H_{\rm
ex}=\frac{J}{2}\sum_{kk^{\prime}}A^{\dagger}_k\left(\begin{array}{cccc}
S_z&&S_-&\\&S_z&&-S_-\\S_+&&-S_z&\\&-S_+&&-S_z\end{array}\right)A_{k^{\prime}},\label{intermag}\\
&H^{\prime}=\frac{V}{2}\sum_{kk^{\prime}}A^{\dagger}_k\left(\begin{array}{cccc}
1&&&\\&-1&&\\&&1&\\&&&-1\end{array}\right)A_{k^{\prime}}.
\label{internonmag}
\end{eqnarray}
where $S_{\pm}=S_x\pm iS_y$. Then the full GF of a $d+id$
superconductor with a classical spin or an ordinary potential is
obtained through the equation of motion method \cite{Shiba1968} of GF by
\begin{eqnarray}
G_{kk^{\prime}}(\omega)=G^0_k(\omega)\delta_{kk^{\prime}}+G^0_k(\omega)t(\omega)G^0_{k^{\prime}}(\omega),
\end{eqnarray}
where $G^0_k(\omega)=1/(\omega-H_{\rm BdG})$ is the free GF of
$d+id$ superconductors, while the T-matrix for a localized spin and
an ordinary potential are separately
\begin{eqnarray}
&& t_{\rm ex}(\omega)=\frac{(JS/2)^2F(\omega)+\left(\begin{array}{cc}
S_z&S_-\tau_z\\S_+\tau_z&-S_z\end{array}\right)}{1-[(JS/2)F(\omega)]^2}
,\label{tmatrixmag}\\
&&
t^{\prime}(\omega)=\frac{H^{\prime}}{1-H^{\prime}F(\omega)}.\label{tmatrixnonmag}
\end{eqnarray}
In the above, the T-matrix depends only on the energy because the
scattering in Eqs. (\ref{intermag}) and (\ref{internonmag}) are both
momentum-independent, the $\tau_{\alpha}$'s are Pauli
matrix in particle-hole space, and we have defined the $F$-function as
\begin{eqnarray}
F(\omega)=\frac{1}{N}\sum_k G^0_k(\omega)={\rm
diag}(F_e\;F_h\;F_e\;F_h) \label{Ffunction}
\end{eqnarray}
with the diagonal elements
\begin{eqnarray}
F_{e(h)}(\omega)=\frac{1}{N}\sum_k\frac{\omega+(-)\varepsilon_k}{\omega^2-\varepsilon_k^2-|\Delta_k|^2}.
\label{Felements}
\end{eqnarray}
It is interesting to notice that this $F$-function in $d+id$ superconductors
is diagonal too as that in the QSH system in HgTe/CdTe quantum wells with $p$-wave symmetry
studied before. \cite{Liu2009} This is because
the off-diagonal terms are momentum integrations of the gap parameter $\Delta_k$,
although $\Delta_k$ has quadratic dependence on momentum,
the integration of real part cancels exactly since the integrand is symmetric with respect to $k_x$
and $k_y$, while the imaginary part vanishes uniformly for that the integrand
is an odd function of $k_x$
and $k_y$. Therefore, the eigen-energies for the LES induced by classical spins
and ordinary potentials are obtained by finding the poles of the
above T-matrix in (\ref{tmatrixmag}) and (\ref{tmatrixnonmag})
respectively for energies in the superconducting gap as
\begin{eqnarray}
&& (JS/2)F_{e(h)}(\omega)=\pm1,\label{resonantmag}\\
&&
(V/2)F_e(\omega)=1,\;{\rm and}\;(V/2)F_h(\omega)=-1,\label{reasonantnonmag}
\end{eqnarray}
which consist of {\it four} conditions when the impurity is
spin-dependent while only {\it two} conditions when the impurity is
spin-independent.
\begin{figure}
\begin{center}
\includegraphics[width=8cm]{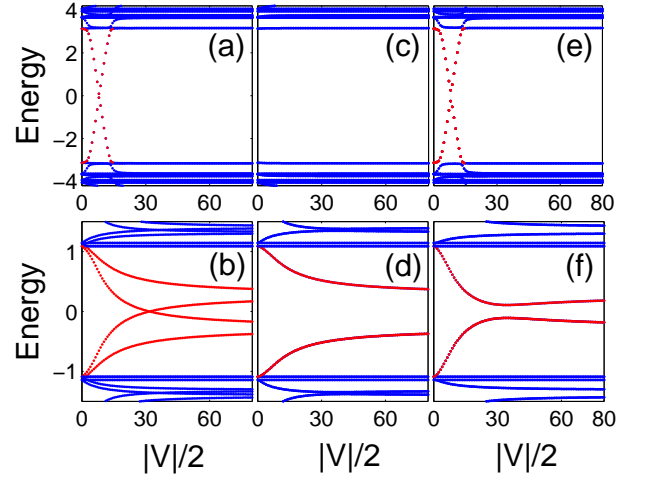}
\caption{(Color online) Real space diagonalization of the total
Hamiltonian $H_{\rm BdG}+H_{\rm ex}$ in (a)-(b) and $H_{\rm BdG}+H^{\prime}$
in (c)-(f) as a function of impurity strength. The
bulk states are represented by blue lines and LES in superconducting gap are
denoted in red lines. Parameters are taken as ${\rm
sgn}(\Delta_{x^2-y^2}/\Delta_{xy})=-1$ in all the figures, $\mu=\pm3$
in topologically nontrivial and trivial phases, and all energies are
measured in units of $t$.
(a)-(b) Topologically trivial and nontrivial phases for
spin-dependent potential. (c)-(d) Topologically trivial and nontrivial phases
with $V>0$ for ordinary potential.
(e)-(f) Topologically trivial and nontrivial phases
with $V<0$ for ordinary potential.} \label{realspacemag}
\end{center}
\end{figure}

To see how these conditions are satisfied for general impurity
strength $J$ and $V$, we numerically examine the momentum
integrations in Eq.(\ref{Felements}) in both topologically
nontrivial and trivial phases which are identified in
Eq.(\ref{skymion}) by constraining the frequency $\omega$ in the
superconducting gap only. Using the tight-binding model where
$\varepsilon_k=t(4-\cos k_x-\cos k_y)-\mu$,
$|\Delta_k|^2=\Delta^2_{x^2-y^2}(\cos k_x-\cos
k_y)^2+\Delta^2_{xy}\sin^2 k_x\sin^2 k_y$, and replacing
$(1/N)\sum_k\rightarrow(1/4\pi^2)\iint dk_xdk_y$, several
representative $F$-functions are plotted versus frequency for
topologically nontrivial phases in Figs.\ref{Ffunctionfig}(a)-(d), and
for topologically trivial phases in Figs.\ref{Ffunctionfig}(e)-(h). A
striking difference of the $F$-functions between topologically
nontrivial and trivial phases is immediately recognized in these
figures. First of all, it is seen that each of the electron band
$F$-function $F_e(\omega)$ and the hole band $F$-function $F_h(\omega)$
winding from $-\infty$ to $\infty$ within the entire range of
$\omega$ in the superconducting gap, and there is a negative ``gap'' between $F_e(\omega)$ and
$F_h(\omega)$ in topologically nontrivial phase [See
Figs.\ref{Ffunctionfig}(a)-(d)]. Therefore, all of the four resonant
conditions in Eq.(\ref{resonantmag}) for classical spins can be
satisfied at any impurity strength $JS$ (for both $J>0$ and $J<0$), and we expect four
persistent LES exist in the superconducting gap. Similarly for
ordinary potential, both of the resonant conditions in
Eq.(\ref{reasonantnonmag}) can also be met with arbitrary $V$ ($>0$ or $<0$), hence
two persistent LES in the superconducting gap is expected when the
impurity is spin-independent. While in a sharp contrast, in
topologically trivial phase, both of the electron band $F$-function
and the hole band $F$-function terminate at some finite value in one
end of number-axis whereas diverge in the other end,
which are separated by a positive ``gap'' [See
Figs.\ref{Ffunctionfig}(e)-(h)]. Consequently, only one condition in
each of $F_e(\omega)$ and $F_h(\omega)$ in Eq.(\ref{resonantmag})
becomes true at small $JS$ but fails for strong enough impurity
strength in the case of classical spins. The case is tricky for
ordinary potentials, which depends on the sign of product of $V$ and $\mu$,
${\rm sgn}(V\mu)$. Since the electron band
$F$-function $F_e(\omega)$ is negative definite and the hole band
$F$-function $F_h(\omega)$ is positive definite for $\omega$ in the
superconducting gap when $\mu<0$ [See Figs.\ref{Ffunctionfig}(e)-(f)], no LES
exist for repulsive interactions but two LES appear for attractive
interactions. And in attractive situation, the two LES finally disappear
at large $V$. Vice versa when $\mu>8$.

As an independent examination of the above statements analyzed solely from the $F$-function
behaviors, we have also directly diagonalized the total Hamiltonian
$H=H_{\rm BdG}+H_{\rm ex}(H^{\prime})$ in real space as a function
of impurity strength $|JS|$ ($|V|$) in Fig.\ref{realspacemag}.
In Figs.\ref{realspacemag}(a) and (b), real space diagonalization
with spin-dependent potential is shown, where we see that in the
topologically trivial phase (up panel), there are indeed two LES
winding through the superconducting gap at small $|JS|$, which disappear
at strong $|JS|$.  Nevertheless, four LES persist in the superconducting
gap at arbitrary impurity strength in
topologically nontrivial phase (lower panel), and the stronger the $|JS|$ is the
deeper the LES are localized in the gap. While in Figs.\ref{realspacemag}(c)-(f),
real space diagonalization
with ordinary potential is presented. We see that in topologically
nontrivial phases for both sign of $V$ [Figs.\ref{realspacemag}(d) and (f)],
two persistent LES exist in the superconducting gap. And in topologically
trivial phases, no LES appear in the superconducting gap
when ${\rm V\mu<0}$ [Fig.\ref{realspacemag}(c)] but two impersistent LES
exist at small $|V|$ when ${\rm V\mu>0}$ [Fig.\ref{realspacemag}(e)].
These pictures support in
perfect to our predictions merely from the $F$-function properties given
in Fig.\ref{Ffunctionfig}. 
In addition, we could expand the $\delta$-function of the impurity
potential and consider an a finite range spin-dependent scattering with $V=V_0\Gamma/(|\textbf{r}-\textbf{r}_{0}|^{2}+\Gamma^{2})$.
For an extended impurity with small $V_0$, the impurity effect is restricted on the impurity site as that in the localized impurity case\cite{Li2012}. For a relative large $V_0$, the impurity scattering is strongest at the impurity site and decay quickly as the lattice spacings increase, which is not important beyond few lattice spacings from the impurity cite\cite{Balatsky2006}. Then, it is expected that similar results could be addressed for a finite range spin-dependent scattering, and detail results should be present in our future work.

\begin{figure}
\begin{center}
\includegraphics[width=8cm]{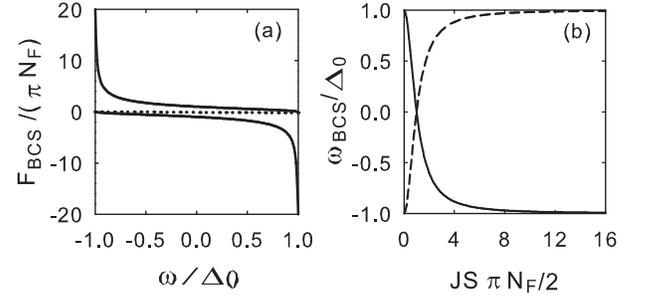}
\caption{(a) $F$-function in BCS superconductors where the electron and hole band $F$-functions terminate
at the same value (see the dotted line) and form a zero ``gap''.
(b) Eigen-energies of two LES induced by classical spins versus
impurity strength in a BCS superconductor.} \label{FShiba}
\end{center}
\end{figure}
For comparison, we now discuss the $F$-functions as well as LES
induced by classical spins and nonmagnetic impurities in BCS superconductors,
which have been studied in early times by H. Shiba. \cite{Shiba1968}
In a BCS superconductor, the corresponding $F$-function is analytically
obtained as $F_{\rm BCS}(\omega)=-\pi N_F((\omega/\Delta_0)\pm1)/\sqrt{1-(\omega/\Delta_0)^2}$,
where $\Delta_0$ is the real gap parameter and $N_F$ is the
density of state at Fermi energy in normal state.
This $F$-function $F_{BdG}(\omega)/(\pi N_F)$ versus energy is
plotted in Fig.\ref{FShiba}(a) for $|\omega/\Delta_0|<1$,
where we see that both of the electron and hole band $F$-functions diverge at
one end while terminate at the same value at the other end.
Therefore, in this sense, a BCS superconductor can be viewed as a zero ``gap''
superconductor compared to the positive ``gap'' topologically trivial and
negative ``gap'' topologically nontrivial phases of a TSC.
For classical spins in a BCS superconductor, there are always two
LES in the superconducting gap at energies $\pm\omega_{BCS}=\pm\Delta_0(1-(JS\pi N_F/2)^2)/(1+(JS\pi N_F/2)^2)$
as shown in Fig.\ref{FShiba}(b), where we see that they behave
critically between the topologically trivial and nontrivial phases
exhibited in Fig.\ref{realspacemag}. The two LES in
BCS superconducting gap persistently tend nearer to the band edges as the impurity strength
goes to infinity but never disappear. While in contrast, the two LES
of the trivial phase in a TSC goes to the band edges as the increase
of impurity strength and then disappear into the bulk at some finite value of $|JS|$ or $|V|$.
For ordinary potential in BCS superconductors, it is shown that
LES never appear because the corresponding T-matrix has no singularities
in the superconducting gap.

The response of a TSC to classical spins and nonmagnetic impurities
compared with that of a conventional BCS superconductor is
summarized in Table. \ref{compare}, where those persistent LES which
survive at arbitrary impurity strength are numbered in bold. This
result provides an effective way to distinguish the topologically
trivial and nontrivial phases of a TSC, and also to tell a
conventional BCS superconductor from a potential TSC, while without
resorting to the surface properties. This is the main result of our
work.
\begin{table}[!h]
\tabcolsep 0pt \caption{The number of LES in the superconducting gap
induced by classical spins and nonmagnetic impurities in a TSC as
well as in a conventional BCS superconductor. The bold numbers
indicate that the corresponding LES exist at arbitrary impurity
strength.} \vspace*{-5pt}
\begin{center}
\def\temptablewidth{0.42\textwidth}
{\rule{\temptablewidth}{1pt}}
\begin{tabular*}{\temptablewidth}{@{\extracolsep{\fill}}ccccccc}
 & trivial-TSC  & nontrivial-TSC & BCS \\   \hline
$JS/2$    &  2   & ${\bf4}$ & ${\bf2}$ \\
$V/2$ & 0, sgn$(V\mu)<0$ & ${\bf 2}$ &  0 \\
& 2, sgn$(V\mu)>0$& &
       \end{tabular*}
       {\rule{\temptablewidth}{1pt}}
       \end{center}
       \label{compare}
       \end{table}

\section{Conclusions and Discussions}
\label{conclusion}
In conclusion, classical spins in generic TIs and TSCs are studied in
an earlier \cite{Liu2009} and the present work. In particular,
an $F$-function is defined in such systems, where the differences of which
from electron and hole band classify BdG Hamiltonian of superconductors into
positive, negative, and zero ``gap'' categories respectively for
topologically trivial and nontrivial phases of a TSC and a BCS superconductor.
The characteristic features of $F$-functions determine directly the
presence or absence of LES, induced by bulk classical spins and
nonmagnetic impurities, in superconducting gap as well as their persistence
with respect to impurity strength. The responses of TSC and BCS superconductors
to bulk classical spins and nonmagnetic impurities are summarized,
where it is shown that the potential
strength of both spin-dependent and ordinary impurities can be used
to tell the topologically nontrivial from the trivial phases of a
TSC, whereas nonmagnetic impurities can be used to distinguish
a TSC from an BCS superconductor. Our results provide an alternative
way to identify topologically insulating and superconducting phases
in experiments without resorting to the surface properties.

\acknowledgements

F. Wang and T. X. Ma are supported by the Research funding for undergraduate
student of BNU, NSFC Grant. No. 11104014,
Research Fund for the Doctoral Program of Higher Education of China
20110003120007, SRF for ROCS (SEM), the Fundamental Research  Funds for the Central Universities in China under 2011CBA00102, and Youth Science Fund Project of BNU. Q. Liu and X. Y. Jiang are
supported by the NSFC (Grant No. 11004212, 11174309, 60877067 and
60938004), and the STCSM (Grant No. 11ZR1443800).


\begin{thebibliography} {s1}
\bibitem{Review} For reviews, see M.Z. Hasan and C.L. Kane, Rev.
Mod. Phys. {\bf 82}, 3045 (2010). X.-L. Qi and S.-C. Zhang, Phys.
Today {\bf 63}, 33 (2010); Rev. Mod. Phys. {\bf 83}, 1057 (2011). M.Z. Hasan and
J.E. Moore, Ann. Review.Condensed Matter Physics {\bf 2}, 55 (2011). Also see the references given
therein.
\bibitem{Liu2009} Qin Liu and Tianxing Ma, Phys. Rev. B {\bf 80},
115216 (2009).
\bibitem{Qi2006} X.-L. Qi, Yong-Shi Wu, and S.-C. Zhang, Phys. Rev. B {\bf 74}, 085308 (2006).
\bibitem{Bernevig2006} B. A. Bernevig, T. L. Hughes, S.-C. Zhang, Science \textbf{314}, 1757 (2006).
\bibitem{Fu2007} L. Fu and C. L. Kane, Phys. Rev. B {\bf 76}, 045302 (2007).
\bibitem{TSC} A. Kitaev, AIP Conf. Proc. {\bf 1134}, 22. X.-L. Qi,
T. L. Hughes, S. Raghu, and S.-C. Zhang, Phys. Rev. Lett. {\bf 102},
187001 (2009). R. Roy, e-print arXiv:0803.2868.
\bibitem{Schnyder2008} A. P. Schnyder, S. Ryu, A. Furusaki, and A. W. W. Ludwig,
Phys. Rev. B {\bf 78}, 195125 (2008).
\bibitem{Mackenzie2003}A. P. Mackenzie, and Y. Maeno, Rev. Mod. Phys. {\bf 75}, 657 (2003).
\bibitem{Lee2009} P. A. Lee, e-print arXiv:0907.2681 (to be published).
\bibitem{Fu2008} L. Fu, and C. L. Kane, Phys. Rev. Lett. {\bf 100},
096407 (2008).
\bibitem{Qi2010} Xiao-Liang Qi, T. L. Hughes, and S.-C. Zhang, Phys. Rev. B {\bf 81},
134508 (2010); X.-L. Qi, T. L. Hughes, and S.-C. Zhang, Phys. Rev. B, {\bf 82},
184516 (2010).
\bibitem{Sau2010} J. D. Sau, R.M. Lutchyn, S. Tewari, and S. Das Sarma, Phys.
Rev. Lett. {\bf 104}, 040502 (2010).
\bibitem{Alicea2010} J. Alicea, Phys. Rev. B {\bf 81}, 125318 (2010).
\bibitem{Lutchyn2010} R. M. Lutchyn, J. D. Sau, and S. Das Sarma, Phys. Rev. Lett. {\bf 105},
077001 (2010).
\bibitem{Brouwer2011}P. W. Brouwer, M. Duckheim, A. Romito, and F. von Oppen, Phys. Rev. B {\bf 84},
144526 (2011).
\bibitem{Mao2012} L. Mao, M. Gong, E. Dumitrescu, S. Tewari, and C. Zhang, Phys. Rev. Lett. in press (2012).
\bibitem{Nayak2008} C. Nayak, S. H. Simon, A. Stern, M. Freedman, and S. D. Sarma,
Rev. Mod. Phys. {\bf 80}, 1083 (2008).
\bibitem{Shiba1968}H. Shiba, Prog. Theor. Phys. {\bf 40}, 435 (1968).
\bibitem{classicalspin} By classical spins, here and all throughout the paper, we refer to the
classcal limit of magnetic impurities where no quantum effect of
spin-flipping is considered. In this sense, the classical spins can
be viewed as a spin-dependent potential, whereas nonmagnetic
impurites are treated as ordinary (spin-independent) potentials.

\bibitem{Read2000} N. Read and D. Green, Phys. Rev. B {\bf 61}, 10267 (2000).
\bibitem{Senthil1999} T. Senthil, J. B. Marston, and Matthew P. A.
Fisher, Phys. Rev. B {\bf 60}, 4245 (1999).
\bibitem{note} TSCs and TIs in one dimension are not included in our discussion.
\bibitem{Sigrist1991} Manfred Sigrist and Kazuo Udea, Rev. Mod.
Phys. {\bf 63}, 239 (1991).

\bibitem{Balian1963}R. Balian and N. R. Werthamer, Phys. Rev. {\bf 131}, 1553 (1963).
\bibitem{Maki1967} K. Maki, Phys. Rev. {\bf 152}, 428 (1967).
\bibitem{Rajaraman} R. Rajaraman, {\it Solitons and Instantons}, North-Halland Publishing Company, 1982.
\bibitem{Li2012} J. Li and C. S. Ting, Phys. Rev. B {\bf 85}, 094520(2012).
\bibitem{Balatsky2006} A. V. Balatsky, I. Vekhter, and J.-X. Zhu, Rev. Mod. Phys. {\bf 78}, 373(2006).

\end{thebibliography}
\end{document}